\documentclass[preprint2]{aastex}
\usepackage{rotating}
\usepackage{graphicx}
\usepackage{grffile}

\begin{document}
\setcounter{secnumdepth}{2}
\title{The TRENDS High-Contrast Imaging Survey. V. \\ Discovery of an Old and Cold Benchmark T-dwarf Orbiting the Nearby G-star HD 19467}
\author{Justin R. Crepp\altaffilmark{1}, John Asher Johnson\altaffilmark{2}, Andrew W. Howard\altaffilmark{3}, Geoffrey W. Marcy\altaffilmark{4}, John Brewer\altaffilmark{5}, Debra A. Fischer\altaffilmark{5}, Jason T. Wright\altaffilmark{6,7}, Howard Isaacson\altaffilmark{4}}
\email{jcrepp@nd.edu} 
\altaffiltext{1}{Department of Physics, University of Notre Dame, 225 Nieuwland Science Hall, Notre Dame, IN, 46556, USA}
\altaffiltext{2}{Harvard Center for Astrophysics, 60 Garden Street, Cambridge, MA 02138, USA} 
\altaffiltext{3}{Institute for Astronomy, University of Hawaii, 2680 Woodlawn Drive, Honolulu, HI 96822}
\altaffiltext{4}{Department of Astronomy, University of California, Berkeley, CA 94720} 
\altaffiltext{5}{Department of Physics, Yale University, New Haven, CT 06511, USA} 
\altaffiltext{6}{Department of Astronomy \& Astrophysics, The Pennsylvania State University, University Park, PA 16802, USA} 
\altaffiltext{7}{Center for Exoplanets and Habitable Worlds, The Pennsylvania State University, University Park, PA 16802, USA}

\begin{abstract}  
The nearby Sun-like star HD~19467 shows a subtle radial velocity (RV) acceleration of $-1.37\pm0.09$ m$\:$s$^{-1}\:$yr$^{-1}$ over an $16.9$ year time baseline (an RV trend), hinting at the existence of a distant orbiting companion. We have obtained high-contrast adaptive optics images of the star using NIRC2 at Keck Observatory and report the direct detection of the body that causes the acceleration. The companion, HD~19467~B, is $\Delta K_s=12.57\pm0.09$ mag fainter than its parent star (contrast ratio of $9.4\times10^{-6}$), has blue colors $J-K_s=-0.36\pm0.14$ ($J-H=-0.29\pm0.15$), and is separated by $\rho=1.653\pm0.004\arcsec$ ($51.1\pm1.0$ AU). Follow-up astrometric measurements obtained over an 1.1 year time baseline demonstrate physical association through common parallactic and proper motion. We calculate a firm lower-limit of $m\geq51.9^{+3.6}_{-4.3}M_J$ for the companion mass from orbital dynamics using a combination of Doppler observations and imaging. We estimate a model-dependent mass of $m=56.7^{+4.6}_{-7.2}M_{Jup}$ from a gyrochronological age of $4.3^{+1.0}_{-1.2}$ Gyr. Isochronal analysis suggests a much older age of $9\pm1$ Gyr, which corresponds to a mass of $m=67.4^{+0.9}_{-1.5}M_J$. HD~19467~B's measured colors and absolute magnitude are consistent with a late T-dwarf [$\approx$T5-T7]. We may infer a low metallicity of [Fe/H]$=-0.15\pm0.04$ for the companion from its G3V parent star. HD~19467~B is the first directly imaged benchmark T-dwarf found orbiting a Sun-like star with a measured RV acceleration.
%
\end{abstract}                                                                                                                                                                                                                                                                                                                                                                                                                                                                                                                                                                                                                                                                                                                                                                                                                                                                                                                                                                                                                                        

\section{INTRODUCTION}\label{sec:intro}
The TRENDS ({\bf T}a{\bf R}getting b{\bf EN}chmark-objects with {\bf D}oppler {\bf S}pectroscopy) high-contrast imaging survey is a dedicated ground-based program that uses adaptive optics and related technologies to directly detect and study faint companions orbiting nearby stars \citep{crepp_12b}. TRENDS differs from other high-contrast campaigns primarily through its selection of targets. Rather than observing nearby young stars, we select older targets that show clear evidence for the existence of a low-mass companion as the result of years of precise radial velocity (RV) measurements. This evidence manifests as an acceleration in the RV time series which we refer to as a Doppler ``trend". The tradeoff between age and {\it a priori} knowledge of an orbiting companion results in a higher detection efficiency at the expense of mass sensitivity. However, recent advances in high-contrast imaging hardware and techniques will soon permit the detection of massive planets (super-Jupiters) around stars as old as $\approx$1 Gyr at infrared wavelengths \citep{crepp_johnson_11,hinkley_11_PASP,macintosh_12,skemer_12}. Operating primarily from Keck Observatory, the TRENDS program is sensitive to brown dwarf companions of any age for essentially all nearby ($d\lesssim50$ pc) targets \citep{montet_13}. 

TRENDS survey discoveries to date include several benchmark high mass ratio binary stars (HD~53665, HD~67017, HD~71881), a triple star system (HD~8375), and a ``Sirius-like" white dwarf companion orbiting HD~114174 \citep{crepp_12b,crepp_13a,crepp_13b}. By connecting the properties of directly imaged companions to that of their primary star (such as metallicity and age), these objects serve as useful test subjects for theoretical models of cool dwarf atmospheres \citep{liu_10}. Further, the combination of Doppler observations and high-contrast imaging constrains the companion mass and orbit, essential information that brown dwarfs discovered in the field or at wide separations by seeing-limited instruments do not provide. In this paper, we report the direct imaging discovery of an old and cold brown dwarf orbiting the nearby G3V star HD~19467 (Table 1). The companion's subtle gravitational influence on HD~19467 was initially noticed as a Doppler acceleration spanning more than a decade. We show that HD~19467~B is almost certainly a T-dwarf, based on its intrinsic brightness and near-infrared colors.

\begin{table}[!ht]
\centerline{
\begin{tabular}{lc}
\hline
\hline
\multicolumn{2}{c}{HD~19467 Properties}     \\
\hline
\hline
right ascension [J2000]            &    03 07 18.57         \\
declination [J2000]                   &    -13 45 42.42        \\
$B$                                           &     7.65                    \\
$V$                                           &     7.00                     \\
$J$                                            &   $5.801\pm0.020$  \\
$H$                                           &   $5.447\pm0.036$   \\
$K_s$                                       &   $5.401\pm0.026$   \\
d [pc]                                         &    $30.86\pm0.60$   \\
$\mu_{\alpha}$ [mas/yr]            &    $-7.81\pm0.63$  \\
$\mu_{\delta}$ [mas/yr]            &   $-260.77\pm0.71$  \\
\hline
Mass [$M_{\odot}$]           &    $0.95\pm0.02$        \\
Radius [$R_{\odot}$]         &     $1.15\pm0.03$       \\
Luminosity [$L_{\odot}$]   &    $1.34\pm0.08$     \\
$\log R'_{HK}$                   &     $-4.98\pm0.01$     \\
Gyro Age [Gyr]                   &    $4.3^{+1.0}_{-1.2}$  \\
SME Age [Gyr]                   &    $9\pm1$                  \\
$\mbox{[Fe/H]}$                 &    $-0.15\pm0.02$      \\
log g [cm $\mbox{s}^{-2}$] &    $4.40\pm0.06$   \\ 
$T_{\rm eff}$ [$K$]             &    $5680\pm40$        \\
Spectral Type                     &      G3V                   \\
v sini   [km/s]                      &    $1.6\pm0.5$            \\
\hline
\end{tabular}}
\caption{(Top) Coordinates, apparent magnitudes, distance, and proper motion of HD~19467. Magnitudes are from the 2-Micron All Sky Survey (2MASS) catalog of point sources \citep{cutri_03,skrutskie_06}. The parallax-based distance from {\it Hipparcos} uses the refined data reduction of \citealt{van_leeuwen_07}. (Bottom) Host star physical properties are estimated from SME using HIRES template spectra and theoretical isochrones \citep{valenti_fischer_05}. We estimate a gyrochronological age based upon empirical relations incorporating $B-V$ color and $R'_{HK}$ value \citep{mamajek_hillenbrand_08}.} 
\label{tab:star_props}
\end{table}

\begin{figure*}[!t]
\begin{center}
\includegraphics[height=3.2in]{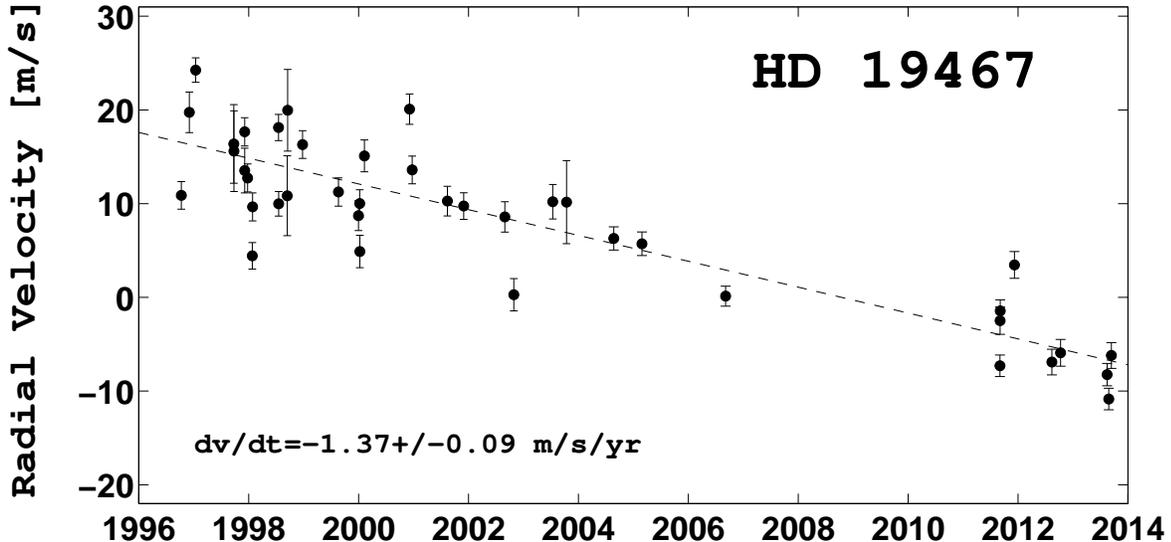} 
\caption{Relative RV measurements of HD~19467. We have directly imaged the substellar companion responsible for the long-term Doppler acceleration.} 
\end{center}\label{fig:image}
\end{figure*} 

\section{OBSERVATIONS}

\subsection{High-Resolution Spectroscopy}

\begin{table*}[!ht]
\centerline{
\begin{tabular}{cccc}
\hline
\hline
Date                      & BJD                         &   RV                    & Uncertainty  \\
 $\mbox{[UT]}$       &      -2,450,000        &  [m~s$^{-1}$]      &  [m~s$^{-1}$]  \\  
\hline
\hline
1996-10-09  & 366.013   & 15.85 &    1.47    \\
1996-12-01  & 418.938   &  24.71   &  2.17   \\
1997-01-13  &  461.838  &  29.22   &  1.30   \\
1997-09-23  &  715.098  &  21.34   &  4.20   \\
1997-09-24  &  716.106  &  20.57  &   4.31   \\
1997-12-04  &  786.842  &  18.50   &  2.38   \\
1997-12-04  & 786.855   & 22.63   &  1.51   \\
1997-12-24  &  806.901  &  17.71  &   1.50   \\
1998-01-24  &  837.743   &  9.40   &  1.41   \\
1998-01-26   &  839.742  &  14.62  &   1.49   \\
1998-07-17  &  1012.120 &   23.09  &   1.40   \\
1998-07-18  &  1013.121  &  14.95   &  1.31   \\
1998-09-13  &  1070.112  &  15.81  &   4.26   \\
1998-09-16   & 1072.980  &  24.93   &  4.36   \\
1998-12-24   & 1171.774  &  21.27  &   1.48   \\
1999-08-19  &  1410.126  &  16.21  &   1.51   \\
1999-12-31  & 1543.844   & 13.69   &  1.60   \\
2000-01-08  &  1551.790  &  14.98  &   1.47   \\
2000-01-09  &  1552.841  &   9.86   &  1.73   \\
2000-02-08  &  1582.730  &  20.06  &   1.70  \\
2000-12-04  &  1882.801  &  25.05   &  1.62   \\
2000-12-22  &  1900.779  &  18.57   &  1.48   \\
2001-08-12  &  2134.078  &  15.24   &  1.58   \\
2001-11-29  &  2242.903  & 14.71   &  1.42   \\
2002-08-29  &  2516.019  &  13.55   &  1.62   \\
2002-10-28  &  2575.896  &   5.25    & 1.72   \\
2003-07-14  &  2835.129  &  15.17   &  1.84  \\
2003-10-13  &  2926.084  &  15.13   &  4.43   \\
     \hline
2004-08-22   & 3240.040   &  6.29   &  1.24   \\
2005-02-26   & 3427.786   &  5.72   &  1.26   \\
2006-09-05   & 3984.035   &  0.13   &  1.07   \\
2011-09-02   & 5807.034   & -7.30   &  1.15   \\
2011-09-03   & 5808.104   & -2.49   &  1.46   \\
2011-09-04   & 5809.087   & -1.44   &  1.18   \\
2011-12-08   & 5903.778   &  3.46   &  1.43   \\
2012-08-12   & 6152.110   & -6.90   &  1.37   \\
2012-10-09   & 6210.014   & -5.92   &  1.43   \\
2013-08-14   & 6519.084   & -8.24   &  1.19   \\
2013-08-25   & 6530.024   & -10.85  &  1.15  \\
2013-09-12   & 6548.034    & -6.20   &  1.38   \\
  \hline
  \hline
\end{tabular}}
\caption{Doppler RV measurements for HD~19467.} 
\vspace{1in}
\end{table*}

\subsubsection{Radial Velocity Measurements}
We obtained RV data for HD~19467 using HIRES at Keck \citep{vogt_94,marcy_butler_92}. First epoch RV observations were acquired on 1996 October 09 UT. Several years of measurements revealed that the star exhibits a persistent acceleration, and that $S_{HK}$ magnetic activity values do not correlate with the RV drift (Fig. 1). Precise Doppler measurements taken over an 16.9 year time frame are listed in Table 2. A horizontal line denotes the location of an RV offset resulting from the Summer 2004 HIRES detector upgrade which we include as a free parameter. A linear fit to the time series yields an acceleration of $-1.37\pm0.09\:\rm{m\:s}^{-1}\:\rm{yr}^{-1}$.


The RV time series also shows significant variations in addition to the systemic acceleration. Fourier analysis based on data acquired through the year 2012 had previously identified a periodic signal at $\approx1.6$ years. However, three more recent observations reveal the $\approx5$ m$\;$s$^{-1}$ signal to be spurious. The level of astrophysical noise (jitter) nominally expected from this type of main-sequence star is $2.4\pm0.4$ m$\;$s$^{-1}$ given its $\log R'_{HK}$ value and $B-V$ color \citep{isaacson_fischer_10}. Using Monte Carlo techniques that randomly scramble measurements in the Doppler time-series, we find that residual RV scatter seen in Fig. 2 (when comparing a linear fit with two-body Keplerian orbital models) results in a false-alarm probability well above the  $\approx1\%$ threshold nominally used for Doppler discoveries \citep{marcy_05}. We consider the additional signal to be spurious and most likely caused by stellar activity, rather than an exoplanet, although further measurements are warranted.

\subsubsection{Star Properties}
Stellar template spectra (non-iodine measurements) were analyzed using the LTE spectral synthesis code {\it Spectroscopy Made Easy} (SME) described in detail in \citet{valenti_96,valenti_fischer_05}. The estimated physical properties of HD~19467 derived from spectral fitting are shown in Table 1. HD~19467 is listed in the SIMBAD database as an G1V star from medium resolution spectroscopy \citep{gray_06}. We find a best fitting spectral type of G3 using higher resolution ($R\approx55,000$) spectroscopy, and a luminosity class of dwarf (V) as we discuss in what follows. 

HD~19467 is a nearby field star not obviously associated with any moving group or cluster. To facilitate our characterization of its companion, we estimate its gyrochronological age using the technique of \citealt{barnes_07}. The stellar rotation period is found empirically to be $P_r=24.9\pm2.5$ days from the measured $\log R'_{HK}$ and $B-V$ values \citep{wright_04}, which corresponding to a gyrochronological age of $4.30^{+0.96}_{-1.23}$ Gyr; this result is based upon updated coefficients that correlate the rotation period and $B-V$ color to age as determined by \citet{mamajek_hillenbrand_08}. Uncertainty in the age is dominated by intrinsic scatter in the empirical relation. 

We have also attempted to calculate a system age using the iterative version of SME that self-consistently incorporates results from the LTE spectral analysis with Yonsei-Yale theoretical isochrones \citep{valenti_fischer_05}. Unfortunately, our code does not converge properly for HD~19467~A. Upon iterating, the age diverges to the end of the grid at 13.7 Gyr. Using only a single iteration, we find a much older age of $9\pm1$ Gyr compared to the gyrochronology method. G3 dwarfs may still reside on the main-sequence at this age, and the low metallicity ([Fe/H]$=-0.15\pm0.04$) of the host star does not suggest youth, but the unusual behavior of the SME iterative code casts doubt on its reliability for this particular source. As such, values listed in Table 1 (age, mass, radius, luminosity) are tabulated using the non-iterative version of SME. We note that HD~19467~B is too faint to cause any substantive spectral contamination. 

Comparing to other Sun-like stars in the solar neighborhood, HD~19467 resides only $\Delta M_V=0.28$ mag above the median Hipparcos-based main-sequence at visible wavelengths \citep{wright_05}. We thus adopt a luminosity class of V. For subsequent analysis, we also adopt the gyrochronological age, noting however that the subsolar metallicity, $[Fe/H]=-0.15\pm0.04$, indicates an age older than the Sun (4.6 Gyr). In Section 3.2, we show that the model-dependent mass of HD~19467~B is still within the brown dwarf regime even for ages up to 10 Gyr. 

\begin{figure}[!t]
\begin{center}
\includegraphics[height=3.2in]{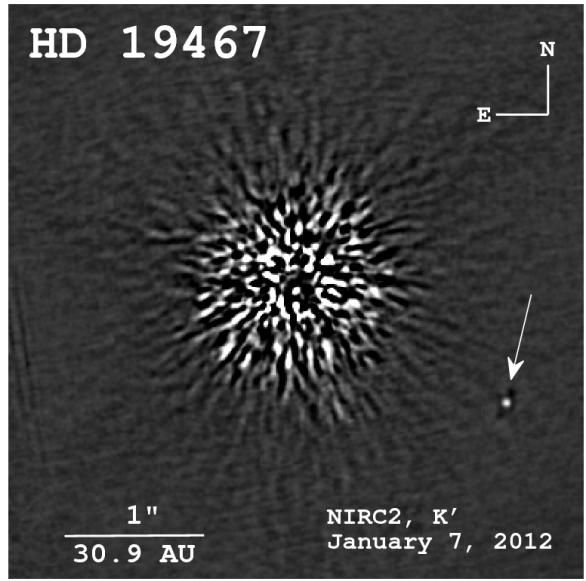} 
\caption{High-contrast image of HD~19467~B taken with NIRC2 AO at Keck Observatory. Stellar speckles have been removed using PSF subtraction. The companion is 100,000 times fainter than its host star in the $K$-band.} 
\end{center}\label{fig:image}
\end{figure} 

\subsection{High-Contrast Imaging}
First-epoch high-contrast images of HD~19467 were acquired with the K' filter on 2011 August 30 UT using NIRC2 (instrument PI: Keith Matthews) and the Keck II AO system \citep{wizinowich_00}. The bright ($K_s=5.401\pm0.026$) star was placed behind the 300 mas diameter coronagraphic spot. We used angular differential imaging (ADI) to enable point-spread-function (PSF) subtraction \citep{marois_06}. Images were processed using the same techniques applied in previous TRENDS discoveries to determine photometric and astrometric quantities (see \citealt{crepp_13b} for details). 


We originally noticed HD~19467~B using raw frames viewed by the NIRC2 graphical user interface, which enables basic data operations such as image subtraction. The companion is fainter than the sky background in K-band under median seeing, but sufficiently separated from the star ($\theta=1.6"$) such that it is detectable by-eye using two subtracted exposures having a small amount of (parallactic) angular diversity. Figure 2 shows a fully processed image of the companion taken on 2012 January 7 UT. HD~19467 was observed at three subsequent epochs to acquire photometric information in complementary filters and assess whether the faint source shares a common parallactic and proper-motion with the star. 

Photometric measurements are summarized in Table 3. HD~19467~B is $\Delta K_s=12.57\pm0.09$ magnitudes fainter than HD~19467~A and has blue colors, $J-K_s=-0.36\pm0.34$ mag, and $J-H=-0.29\pm0.15$ mag. Our astrometric observations consist of four epochs taken over an 1.1 year baseline (Table~4). The proper-motion of HD~19467 is $260.9\pm0.7$ mas $\mbox{yr}^{-1}$. Meanwhile, the size of a NIRC2 pixel is $9.963\pm0.006$ mas as projected onto the sky \citep{ghez_08}. Comparing our relative astrometry measurements to the expected motion (vector sum of parallax and proper-motion) of an unrelated distant background source (i.e., null hypothesis), we find that HD~19467~B is clearly associated with HD~19467~A (Fig. 3). HD~19467~B has a projected separation of $51.1\pm1.0$ AU (2012-10-04 UT) and appears to exhibit clockwise orbital motion at a level of $22\pm6$ mas yr$^{-1}$.

\begin{table}[!ht]
\centerline{
\begin{tabular}{lc}
\hline
\hline
\multicolumn{2}{c}{Imaged Companion: HD~19467~B}     \\
\hline
\hline
$\Delta J$                    &      $11.81\pm0.10$         \\
$\Delta H$                   &      $12.46\pm0.10$         \\
$\Delta K_s$                &      $12.57\pm0.09$        \\
$J$                               &     $17.61\pm0.11$         \\
$H$                              &     $17.90\pm0.11$          \\
$K_s$                           &    $17.97\pm0.09$          \\
$M_J$                           &    $15.16\pm0.12$              \\
$M_H$                          &    $15.45\pm0.12$               \\
$M_{K_s}$                    &     $15.52\pm0.10$              \\
$m_{\rm dyn}$ [$M_J$]        &    $>51.9^{+3.6}_{-4.3}$  \\
$m_{\rm model}$ [$M_J$]    &    $56.7^{+4.6}_{-7.2}$  \\
\hline
\end{tabular}}
\caption{Photometric results and companion physical properties. The mass constraint (lower-limit) from dynamics ($m_{\rm dyn}$) using RV and imaging measurements is consistent with the model-dependent mass estimate from photometry ($m_{\rm model}$). The listed model-dependent mass is based upon the gyrochronological age of the primary star.} 
\label{tab:comp_props}
\end{table}

\section{HD~19467~B PHYSICAL PROPERTIES}
\subsection{Dynamical Mass}
The measured RV acceleration may be combined with the companion projected separation to determine a lower-limit to its mass using dynamics \citep{torres_99,liu_02}. The straight-line fit to the RV time series of $-1.37\pm0.09$ m$\:$s$^{-1}\:$yr$^{-1}$ results in a minimum mass of $m\geq51.9^{+3.6}_{-4.3}M_J$, consistent with a non-Hydrogen-fusing object for near-edge-on orbits (Table 3). It will be possible to place an upper-limit on the companion mass and constrain the six orbital elements when the RV's and astrometry both show curvature. 

\subsection{Model-Dependent Mass Estimate from Photometry}
The absolute magnitude of HD~19467~B (Table 3) is found using our measured magnitude difference and precise Hipparcos parallax of $32.40\pm0.62$ mas \citep{van_leeuwen_07}. A value of $M_{K_s}=15.52\pm0.10$ corresponds to a mass of $56.7^{+4.6}_{-7.2}M_J$ and effective temperature $T_{\rm eff}=1050\pm40$ K according to \citealt{baraffe_03} (COND) evolutionary models using our derived gyrochronological age. Assuming the host star is instead $9\pm1$ Gyr old, as indicated by an isochronal analysis [see notes from $\S$2.1], we find a model-dependent mass still within the brown dwarf regime, $m=67.4^{+0.9}_{-1.5}M_J$. 

Figure 4 shows a plot of the companion measured absolute magnitude and $J-K$ color compared to T-dwarfs characterized using spectroscopy from \citealt{leggett_10}. HD~19467~B's brightness and spectral energy distribution are consistent with an $\approx$T5-T7 dwarf, affirming the interpretation of a cold substellar object. The $J-H=-0.29\pm0.16$ color is also consistent with a T-dwarf classification of $\approx$T5-T8 \citep{cushing_11,kirkpatrick_13,dupuy_liu_12}. Moderate resolution spectroscopy obtained with an integral field spectrograph will establish a more robust spectral-type designation [e.g., \citealt{hinkley_11_PASP}]. 

\subsection{Physical Separation and Period Range}   
We can constrain the system orbital separation and period using the instantaneous RV acceleration and the mass estimate of HD~19467~B from photometry \citep{howard_10}. Such analysis indicates that HD~19467~B is presently $72.5\pm3.3$ AU from its host star, (self-consistently) further than the projected separation ($51.1\pm1.0$ AU). Assuming the true anomaly is near extreme values, i.e., apastron or periastron, we find that the orbital period lies between 320--1900 years for eccentricities between $0 \leq e \leq0.5$ using a host star mass of $M_*=0.95\pm0.02M_{\odot}$ (Table 1). Using the Vis-viva equation, which relates orbit velocity to the instantaneous separation and semimajor axis, we find that the maximum projected sky-motion is $28.0\pm0.9$ mas yr$^{-1}$ for the same eccentricity range, consistent with our observation that the astrometric position of HD~19467~B has changed by $22\pm6$ mas yr$^{-1}$. Our numerical simulations suggest that a unique solution for the orbit and mass of HD~19467~B may be determined as early as $\approx$15 years from now with sufficient observing cadence and continued Doppler and imaging monitoring \citep{crepp_12AAS}. Improved astrometric precision can of course facilitate a more rapid (shortened) orbit characterization timescale.

\begin{table*}[t]
\centerline{
\begin{tabular}{lccccccc}
\hline
\hline
Date [UT]    &   JD-2,450,000  &   Filter  &   $\Delta t$ [min.]   &  $\Delta \pi$ [$^{\circ}$]   &    $\rho$ [mas]      &    P. A. [$^{\circ}$]    &     Proj. Sep. [AU]    \\
\hline
\hline        
2011-08-30      &   5,804.1        &      $K'$                &    20.8             &     12.3       &   $1662.7\pm4.9$     &   $243.14\pm0.19$    &   $51.4\pm1.0$     \\
2012-01-07      &   5,933.8        &      $J$                  &   5.0                &   3.6           &            -----                    &              -----                &        -----    \\
2012-01-07      &   5,933.8        &      $H$                 &    7.5               &   4.4           &   $1665.7\pm7.0$     &   $242.25\pm0.26$    &   $51.4\pm1.0$    \\
2012-01-07      &   5,933.8        &      $K'$                &      7.5              &   4.6          &   $1657.3\pm7.2$     &   $242.39\pm0.38$    &   $51.2\pm1.0$    \\
2012-08-26      &   6,166.1        &      $K'$                &      8.0             &     4.5         &   $1661.8\pm4.4$     &   $242.19\pm0.15$    &   $51.3\pm1.0$    \\
2012-10-04      &   6,205.0        &      $K_s$             &      10.0           &    5.4          &  $1653.1\pm4.1$     &   $242.13\pm0.14$    &   $51.1\pm1.0$    \\
\hline
\hline
\end{tabular}}
\caption{Summary of high-contrast imaging observations, including integration time ($\Delta t$) and parallactic angle rotation ($\Delta \pi$), and resulting astrometric measurements. The primary star was not visible through the coronagraphic mask on 2012 January 07 UT in J-band due to the lower Strehl ratio. Proper-motion and parallactic motion analysis uses a weighted average for H, K' astrometry from this epoch.}
\label{tab:astrometry}
\end{table*}

\begin{figure*}[!t]
\begin{center}
\includegraphics[height=3.4in]{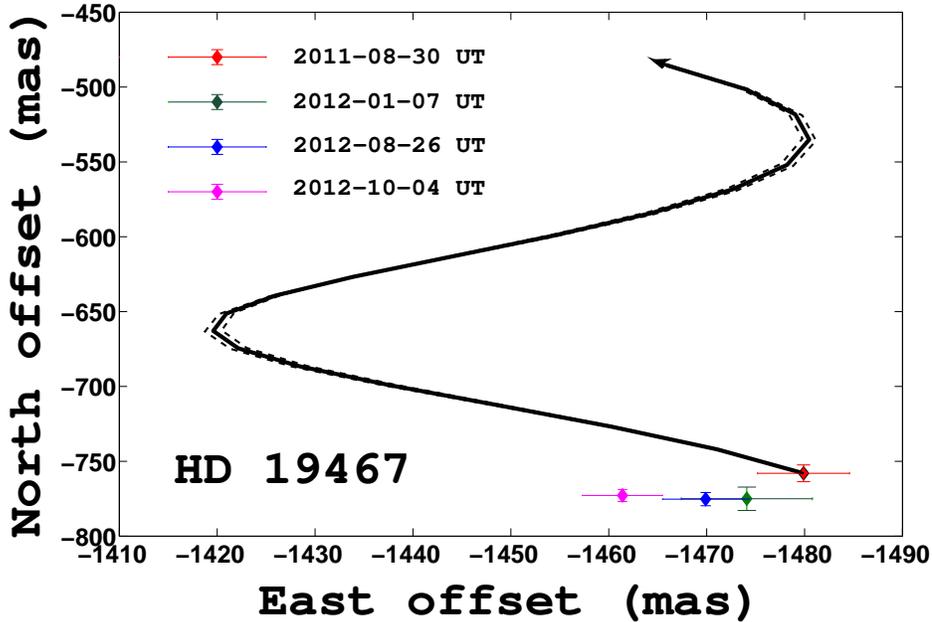} 
\caption{Relative astrometry results showing angular offset of the companion from the primary star at each epoch. The location of HD~19467~A defines the coordinate system origin. The solid curve shows the path that an infinitely distant background object would traverse from 2011 August 30 UT through 2012 October 04 UT. Dashed curves indicate uncertainty in the system proper motion and parallax. HD 19467~B is associated with its parent star ($42\sigma$). We detect systemic orbital motion of $22\pm6$ mas yr$^{-1}$ in a clockwise direction.} 
\end{center}\label{fig:astrometry}
\end{figure*} 

\begin{figure}[!t]
\begin{center}
\includegraphics[height=4.6in]{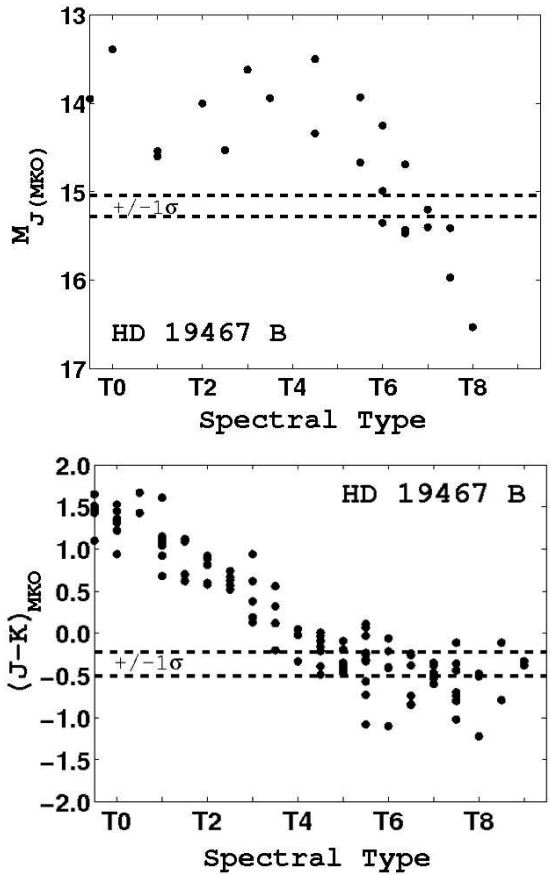} 
\caption{HD~19467~B absolute J-band magnitude and J-K color (dashed lines) compared to field T-dwarfs (filled circles) \citet{leggett_10}. HD~19467~B appears to be consistent with an $\approx$T5-T7 dwarf.} 
\end{center}\label{fig:image}
\end{figure} 

\section{SUMMARY AND DISCUSSION}
We have acquired 40 precise Doppler observations of a nearby ($d=30.86\pm0.60$ pc) G3V star, HD~19467, over an 16.9 year time baseline. A long-term RV drift of $-1.37\pm0.09$ m$\:$s$^{-1}\:$yr$^{-1}$ indicates a distant companion with substellar minimum mass. Follow-up AO observations acquired with NIRC2 at Keck as part of the TRENDS high-contrast imaging program detect the companion responsible for the Doppler acceleration in four different filters and epochs. Astrometric analysis demonstrates unambiguous association of the source with its parent star over a 1.1 year time frame. We detect orbital motion in a clockwise direction at $3.7\sigma$.

HD~19467~B is intrinsically faint, with a contrast ratio ($\Delta K_s=12.57\pm0.09$ mag) comparable to the planets orbiting HR~8799 \citep{marois_10}. Placing the object on an H-R diagram, we find that its absolute magnitude ($M_J=15.16\pm0.12$) and blue colors ($J-K_s=-0.34\pm0.14$, $J-H=-0.29\pm0.16$) are most consistent with well-characterized field brown dwarfs in the $\approx$T5-T7 range \citep{leggett_10}. Theoretical evolutionary models indicate a mass of $m=56.7^{+4.6}_{-7.2}M_{Jup}$ using a gyrochronological age of $4.3^{+1.0}_{-1.2}$ Gyr \citep{baraffe_03,mamajek_hillenbrand_08}. Isochronal analysis suggests a much older age of $9\pm1$ Gyr, resulting in a mass of $m=67.4^{+0.9}_{-1.5}M_J$. Both photometric mass estimates are consistent with the $m\geq51.9^{+3.6}_{-4.3}M_J$ lower-limit derived using dynamics. Assuming a common origin between host star and companion, we may infer a low metallicity for HD~19467~B of [Fe/H]$=-0.15\pm0.04$. Given that the primary star is slightly evolved, residing $\Delta M_V=0.28$ mag above the median Hipparcos main-sequence, a subsolar metallicity implies an age older than that of the Sun ($4.6$ Gyr).

Substellar benchmark objects for which it is possible to simultaneously constrain the mass, age, and chemical composition are scarce but extremely valuable for calibrating theoretical atmospheric models and theoretical evolutionary models \citep{potter_02}. Many nearby T(Y)-dwarfs have been discovered as field objects by surveys that scan large regions of the sky at (red) optical, near-infrared, and mid-infrared wavelengths \citep{cushing_11,bihain_13,liu_13}. A fraction of these objects are members of  multiple systems, but their projected physical separations are large, since the bright glare of host stars often precludes the detection of ultracold dwarfs orbiting in close proximity. Thus, the age of field brown dwarfs is generally highly uncertain and their utility for calibrating theoretical models limited. Of the coldest directly imaged T-dwarfs orbiting Sun-like stars [see \citealt{liu_11} for a list], HD~19467~B is the first with a measured Doppler acceleration, and so will be among the first to have a dynamically measured mass. 

HD~19467~B appears to be a warmer version of GJ~758~B \citep{thalmann_09,janson_11}, except with much bluer colors. Both companions are old T-dwarfs orbiting nearby, well-characterized, main-sequence G- stars with precise parallax measurements. The difference in colors may be indicative of different cloud structures, atmospheric chemistry, or surface gravity \citep{liu_13}. HD~19467~B also represents a T-dwarf analogue to the directly imaged L4 companion to HR~7672~A \citep{liu_02}, which likewise has numerous Doppler measurements that have recently yielded a precise dynamical mass and orbit solution \citep{crepp_12a}. With an apparent magnitude of $J=17.61\pm0.11$ and angular separation of $1.653"\pm0.004"$, it should be possible to obtain high resolution, high signal-to-noise ratio near-infrared spectra of HD~19467~B comparable in quality to that obtained for the outer-planets orbiting HR~8799 \citep{bowler_10,konopacky_13}. Thus, HD~19467~B is an important benchmark object that will complement our understanding of low temperature dwarfs by exploring regions of parameter space corresponding to old age and subsolar metallicity as substellar objects evolve across the H-R diagram in time.  

\section{ACKNOWLEDGEMENTS}
The TRENDS high-contrast imaging program is supported by NASA Origins of Solar Systems grant NNX13AB03G. JAJ is supported by generous grants from the David and Lucile Packard Foundation and the Alfred P. Sloan Foundation. This research has made use of the SIMBAD database, operated at CDS, Strasbourg, France. Data presented herein were obtained at the W.M. Keck Observatory, which is operated as a scientific partnership among the California Institute of Technology, the University of California and the National Aeronautics and Space Administration. The Observatory was made possible by the generous financial support of the W.M. Keck Foundation. The Center for Exoplanets and Habitable Worlds is supported by the Pennsylvania State University, the Eberly College of Science, and the Pennsylvania Space Grant Consortium.

\begin{small}
\bibliographystyle{jtb}
\bibliography{ms.bib}
\end{small}

\end{document}